# Complex structured light generation using printed liquid crystal droplets

Xuke Qiu*, Runchen Zhang, Yifei Ma, Zimo Zhao, Zipei Song, Alva C. J. Orr, Mengmeng Li, Waqas Kamal, Jinge Guo, Alfonso A. Castrejón-pita, Steve J. Elston, Stephen M. Morris*, and Chao He*

*Department of Engineering Science, University of Oxford, Parks Road, Oxford, OX1 3PJ, UK*

E-mail: xuke.qiu@eng.ox.ac.uk; stephen.morris@eng.ox.ac.uk; chao.he@eng.ox.ac.uk

Inkjet-printed liquid crystal (LC) droplets exhibit an intricate spatially-varying birefringence due to their complex internal director configuration. While such anisotropy is often viewed as a drawback when LC droplets are used as microlenses, here we leverage this remarkable birefringence property to generate complex structured light. Through a selection of the alignment layer, and by varying the chiral pitch, we create three distinct droplet types with tailored intrinsic director configurations, each exhibiting a unique birefringence distribution for structured light beam generation. We show that these printed LC droplets can generate beams that exhibit skyrmionic structures carrying two units of orbital angular momentum, beams that contain azimuthal/radial polarized fields, and beams with polarization singularities. Our method enables new possibilities for using LC droplet technology to engineer sophisticated optical beam patterns.

**Introduction**

Structured light refers to electromagnetic fields that possess specific spatial structures, such as in phase and/or polarization[1]. Adhering to this classification, various complex beams with non-uniform polarization distributions across the transverse plane - often referred to as vector beams - have emerged as a crucial class of structured light fields, playing essential roles in modern optics, including quantum information processing, optical trapping and manipulation, and high-density optical data storage [2-8]. Additionally, light with helical phase fronts, known as vortex beams, carry orbital angular momentum (OAM) [9-11] and have proven to be of importance in modern optical communications [12, 13], super-resolution microscopy [14, 15], and optical tweezing [16-18]. When both polarization and phase structures are combined, they form vector vortex beams (VVBs) [19-29], which further expands the capabilities of structured light.

Liquid crystal (LC) spatial light modulators (SLMs) are widely used to generate VVBs through the manipulation of both the phase and polarization of light [30-39]. However, achieving light modulation with spatially varying phase and polarization structures often requires complex optical configurations with multiple SLMs [30, 40], making it both a costly and complicated approach. Employing passive components to generate VVBs offers an attractive alternative when cost or system complexity is a concern. Such passive solutions - demonstrated using Q-plates [41, 42], metasurfaces [24, 43, 44], glass cones [29], and GRIN lenses [20-22, 45-47] - provide advantages such as high precision and design flexibility. However, they often involve complex fabrication processes, high production costs, or bulky optical arrangements, which may limit their practical scalability.

Micro-droplet printing technology has gained significant attention for its ability to precisely pattern complex functional fluids at the microscale on a range of different substrates [48, 49]. This approach enables rapid and efficient deposition of "inks" with wide-ranging characteristics, significantly enhancing production throughput. Consequently, this technique has been readily employed to print a range of materials, including organic light-emitting substances, particle suspensions, conductive polymers, and metalloids, broadening its industrial and technological applications [50-54]. Regarding the printing of LC materials, the inherent spatially varying birefringence, arising from unique internal director configuration, has been somewhat less explored; this characteristic is traditionally seen as a limitation when creating functional microlenses [55].

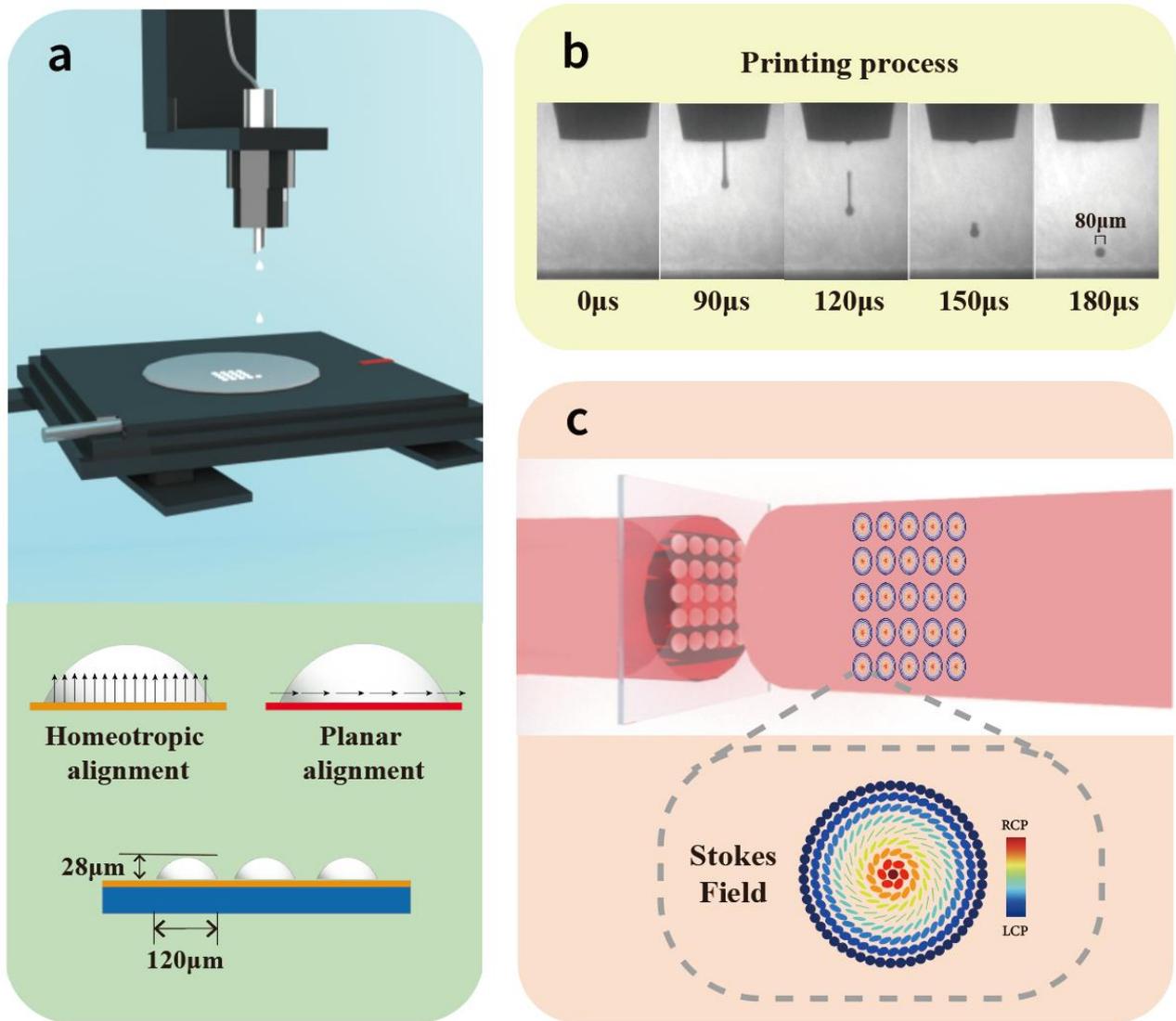

**Figure 1 Schematic representation and visualization of the LC droplet array printing process and the modulation on the polarization characteristics.** (a) Schematic of the inkjet-printing process used to deposit LC droplets onto glass substrates with either homeotropic or planar alignment layers. Each printed droplet measures about 28 μm in height and 120 μm in diameter. (b) Time-lapse shadowgraph images showing the transit of the LC droplets in air at time intervals of 30 μs following ejection from the nozzle. (c) Visualization of light modulation by the printed LC droplet array. Each droplet acts as a birefringent element, altering the incident light's vectorial properties. The resulting Stokes field illustrates the spatial distribution of the polarization states. The color bar, presented using a jet colormap, represents the degree of ellipticity, with color transitions indicating a shift from left-handed to right-handed circular polarization.

In this study, we assess and harness the spontaneous formation of complex internal director configurations within printed nematic and chiral nematic droplets to generate diverse structured light fields. Through use of micro-droplet printing, we deposit LC droplets onto substrates treated with different alignment layers. The resulting structured light generation that is observed includes optical beams exhibiting skyrmionic structures with two units of OAM, beams with azimuthal/radial polarization patterns, and beams with polarization singularities. The approach presented herein provides a simple, compact way to generate complex structured light using LC droplets, paving the way for new LC droplet designs and expanded applications in optical manipulation and information processing.

**Results**

To understand the structured light generation capability of printed LC droplets with intricate internal director profiles, we analyzed the optical properties of three different types of droplets: (1) nematic LC on a homeotropic alignment layer, (2) nematic LC on a planar alignment layer, and (3) a long-pitch chiral nematic LC on a planar alignment layer. For droplets printed onto planar and homeotropic alignment layers, the LC director aligned parallel and perpendicular to the substrate plane at the glass substrate/LC interface respectively. Regardless of the alignment choice, the air/LC interface typically favors a homeotropic alignment of the LC director. Each type of droplet exhibits different birefringence distributions, inducing different polarization and phase modulation characteristics.

In our experiments, we employed a drop-on-demand (DoD) inkjet printing process to deposit LC droplets onto glass substrates coated with either homeotropic or planar alignment layers (**Fig.1a**). Each droplet is printed within a timeframe of 200 μ$s$, with an in-flight diameter of approximately 80 μm (**Fig.1b**). Upon deposition, the droplets typically measure about 28 μm in height and 120 μm in diameter (see **Methods**). **Figure 1c** shows the polarization field from an array of nematic LC droplets, demonstrating that certain droplets can generate a full Poincaré beam [37, 56, 57] (as will be discussed in a later section).

a) **Printed nematic LC on homeotropic alignment layer**

A full Poincaré beam, as a type of structured light field, maps every polarization state on the Poincaré sphere onto specific points within its transverse plane. Among these, the skyrmionic beam [58] stands out as a special case, distinguished by its intricate topological textures in polarization that arise from spatially varying polarization states [47, 59-64]. This section investigates the properties of a nematic LC (E7) droplet deposited on a homeotropic alignment layer, which generates a skyrmionic beam carrying two units of OAM when illuminated with circularly polarized light.

To analyze the optical properties of a printed droplet of the nematic LC onto a homeotropic alignment layer, we first viewed the droplet on a polarizing optical microscope (POM), as shown in **Fig.2a(i)**. The droplet displays the typical two dark brushes that intersect at the center, corresponding to the orientation of the polarizers, where the dark regions denote areas where the LC director is mostly vertical but is slightly tilted toward the polarizer. The bright regions correspond to areas where the local polarization state of transmitted light differs from the extinction direction of the crossed polarizers, resulting in partial transmission [65]. These observations reveal the intrinsic spatial modulation of polarization enabled by the droplet configuration.

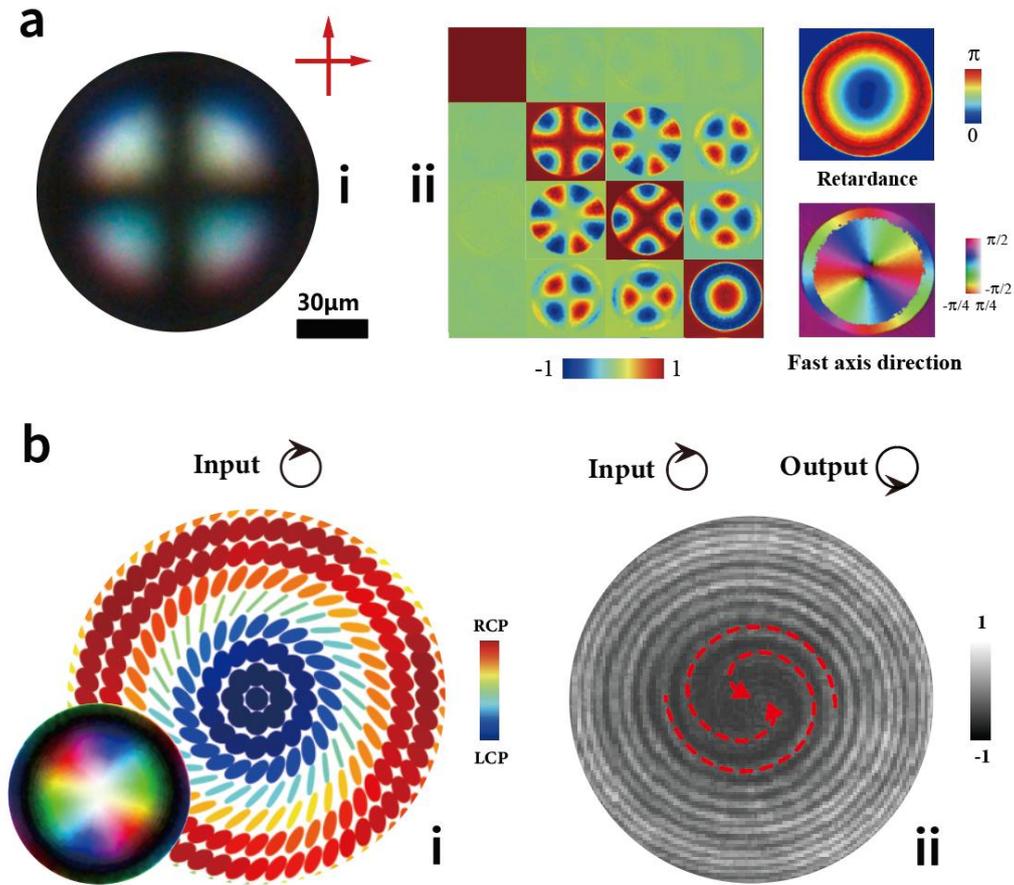

**Figure 2 Experimental results demonstrating the modulation of light by a printed nematic LC droplet on a glass substrate coated with a homeotropic alignment layer.** a(i) POM image of the printed droplet (diameter = 122 μm, height = 29 $\mu m$, contact angle ≈ 32.2°, temperature = 23℃). The single-headed red arrows represent the orientations of the two polarizers. a(ii) The MM characterizes the droplet's full optical response. The MM decomposition reveals the effective linear retardance and fast axis direction of the droplet. The retardance profile shows concentric rings, while the fast axis direction exhibits a radial symmetry, both indicating a vertical alignment with a gentle tendency for the LC director to align radially within the droplet. The fast axis was characterized using the Hue-Saturation-Luminance model. The retardance distribution is represented using the color bar, with values ranging from 0 to $\pi$. b(i) The resulting polarization pattern demonstrates a full Poincaré beam (also exhibiting skyrmionic structure), showcasing a continuous transition of polarization states. The different colors in the pattern represent varying polarization states, with blue indicating left circular polarization and red indicating right circular polarization. b(ii) The interference pattern. The spiral wavefront pattern is clearly visible, and the red dashed lines depict the shape of a light beam with two units of OAM.

The corresponding Mueller matrix (MM) (see **Fig.2a(ii)**) shows the droplet's complete vectorial optical properties. We utilized an MM imaging polarimeter [66] based on the dual rotating waveplates method to measure the LC droplets. Then, by using the MM polarimetry decomposition (MMPD) method [67-81], detailed profile of the retardance and fast axis orientation of the droplet could be obtained, as shown in the two subplots on the right side of the figure. Since the droplet is a continuous birefringent medium, we use an elliptical retarder model to characterize it as a whole, rather than decomposing the system into separate circular and linear retarders. This approach better reflects its non-layered structure and captures the overall polarization modulation along the light propagation path. The retardance profile exhibits a series of concentric rings, with retardance values increasing from the center outward, reaching a peak near the droplet's edge, and then slightly decreasing. Meanwhile, the fast axis direction profile shows a radial symmetry, with the fast axis orientation varying smoothly as a function of angular position around the center. These patterns reveal a clear radial alignment of the LC director within the droplet and are in good agreement with results presented in previous work [11]. The apparent abrupt changes in fast-axis orientation arise from the arctan calculation in MMPD, whose limited range $(-90°, 90°)$ produces artificial jumps of $180°$ when the smoothly varying LC director orientation crosses this boundary.

Using data from both the retardance and fast axis orientation maps, we can predict that the generated light field encompasses all possible polarization states, forming a full Poincaré beam. **Figure 2b** validates the polarization pattern generated by a droplet when illuminated with a right circularly polarized input light. The resulting full Poincaré beam shows a smooth transition of polarization states across the beam's cross-section, exhibiting an intrinsic pattern which is recognized as a Stokes optical skyrmion [47, 61]. Moving radially outward, the polarization transitions from left-handed circular at the center to elliptical, gradually evolving into a linear state at positions corresponding to the Poincaré sphere's equator. Beyond this, the polarization shifts back to circular with reversed handedness, culminating in a right-handed circular state at the beam's edges. At the outermost region, where the retardance reaches its maximum, it then decreases, causing the polarization state to gradually shift back from circular polarization toward elliptical polarization. Despite this, the generated beam still maintains the key characteristics of a full Poincaré beam.

The results presented in **Fig.2a(ii)** also indicate that the effective linear fast axis direction rotates twice around the azimuthal angle, implying the presence of two units of OAM when illuminated with circularly polarized light [46]. To validate this finding, we use a Mach-Zehnder interferometer (see **Supplementary Note 1**) to analyze the phase shift induced by the droplet's internal structure. When placed within a Mach-Zehnder interferometer, the right-handed circularly polarized light from the polarization state generator (PSG) interacts with the droplet, and the left-handed circularly polarized light is filtered by the polarization state analyzer (PSA). The resulting interference pattern, as illustrated in **Fig.2b(ii)**, is characterized by two distinct red arrows. These arrows are indicative that the transmitted light carries two units of OAM, where the number of spiral arms in such an interference pattern directly corresponds to the OAM order of the light (see **Supplementary Note 2**). Typically, spiral interference patterns result from combining a vortex phase and a spherical phase difference between two beams. Here, although the reference arm consists of a plane wave, the observed spiral indicates that the experimental arm itself carried both vortex and spherical-like phase components. This observation implies that the droplet functions as both a 2-OAM generator and a lens with inherent aberration.

### b) Printed nematic LC on Planar alignment layer

It is found that the nematic LC droplet on a planar alignment layer can be used to generate radially/azimuthally polarized light beams. These beams are particularly useful for stimulated emission depletion microscopy and other super-resolution imaging techniques, which can increase resolution beyond the diffraction limit [82-84]. Using a POM and a MM polarimeter as previously described, we elucidated the droplet's structure, as shown in **Fig.3a(i)** and **a(ii)**, respectively. The POM image reveals shadowed regions in the vertical direction and increasing brightness horizontally. This indicates that the director within the droplet predominantly tilts vertically. The MM polarimeter maps in **Fig.3a(ii)**, obtained through MMPD analysis, show that the retardance is symmetric about the horizontal division line, with opposite values on either side.

Having established the polarization-modulating capability of the LC droplet using MMPD, we next examined how the droplet influences both polarization and optical phase for different incident polarization conditions. Specifically, using vertically and horizontally aligned input light, we produced azimuthal and radial polarization patterns, respectively, as shown in **Fig.3b(i)** and **Fig.3c(i)**. To further understand the droplet's modulation of the light field, we conducted an examination of its phase characteristics using the interferometric measurement system described previously (see **Methods**). By illuminating the droplet with either horizontally or vertically polarized light and subsequently filtering the output with a circular polarizer (PSA), we obtained the two images shown in **Fig.3b(ii)** and **Fig.3c(ii)**. Both images exhibit ring-like patterns with fringe discontinuities, but they differ in the size of the phase shift. In one case, a near-$\pi$ shift leads to prominent breaks, while in the other, the shift is minimal, resulting in nearly continuous fringes.

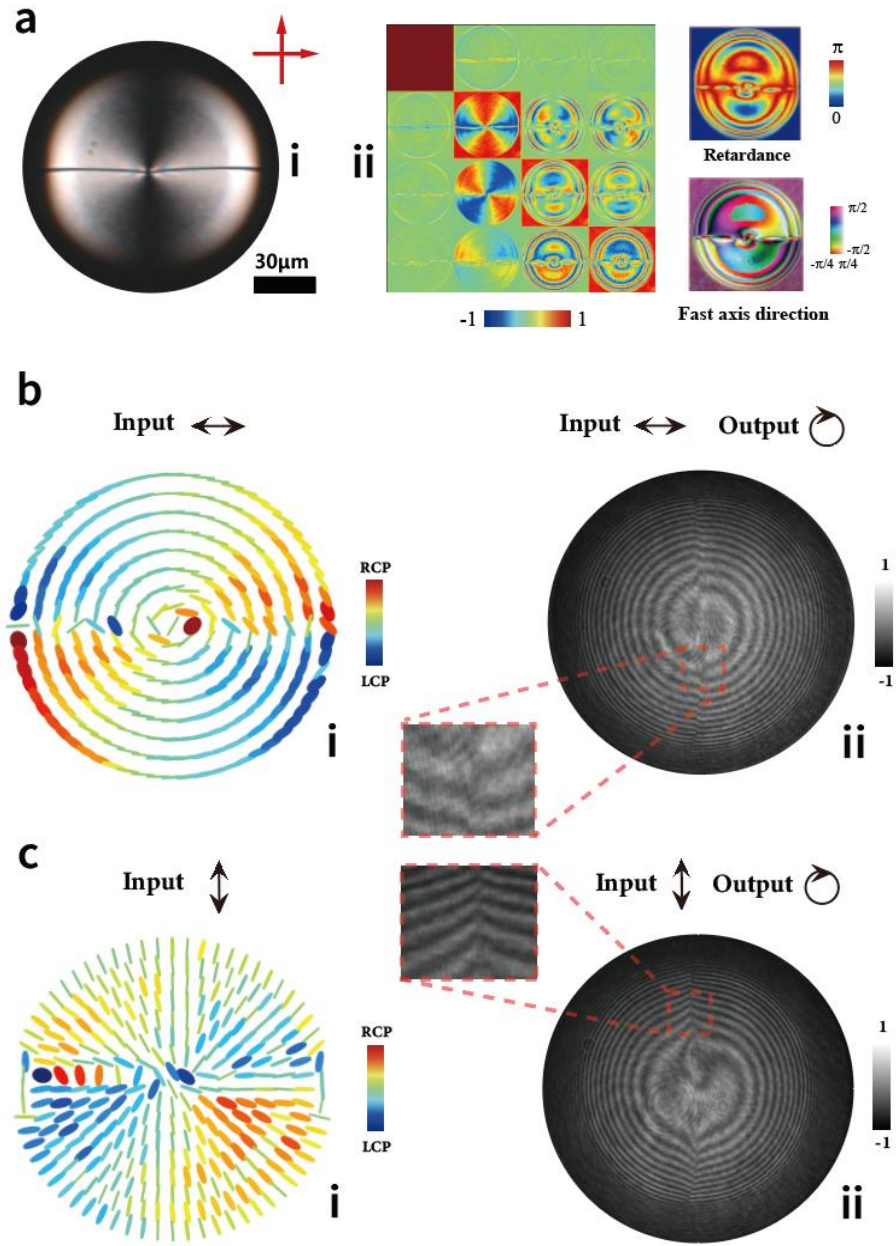

**Figure 3 Experimental results demonstrate the modulation of light by a printed nematic LC droplet (E7) on a glass substrate coated with a planar alignment layer.** a(i) POM image of the printed droplet. The single-headed red arrows represent the orientations of the polarizers. The characteristics of the droplets are as follows: diameter = 125 μm, height = 28 μm, and contact angle = 32.1°. a(ii) The MM components characterizing the droplet's full optical response. The decomposition of the MM provides the retardance and fast axis direction profiles, showing how the droplet alters the polarization. (b) The polarization patterns and the interference patterns when illuminated with horizontally polarized light. (c) The polarization patterns and the interference patterns when illuminated with vertically polarized light. Measurements were carried out at 23°C.

A possible explanation for the observed interference patterns is the influence of the alignment layer beneath the droplet. The planar alignment layer's rubbing direction, which is perpendicular to the disclination line shown in **Fig.3a(i)**, ensures uniform alignment of the LC director near the substrate. As light propagates through the droplet, its phase accumulates from local changes in refractive index that depend on the LC director orientation. On either side of the disclination line, the director is mirror-symmetric but oppositely tilted, leading to different effective retardance values and a phase difference that produces discontinuous fringes. When the incident light is polarized parallel to the rubbing direction, the light first encounters the extraordinary refractive index and then gradually the ordinary refractive index as the director reorients near the droplet's curved surface. Because the director tilt is mirror-symmetric but opposite on either side of the disclination line, the rate of this transition differs between the two regions. This results in a phase difference that is not smoothly varying across the disclination line, leading to discontinuities in the interference fringes. In contrast, with incident light polarized perpendicular to the rubbing direction, the light experiences a uniform ordinary refractive index, yielding nearly equal optical path lengths and continuous interference fringes.

### c) Printed long-pitch chiral nematic LC on a planar alignment layer

In the final section, we demonstrate that LC droplets can be tuned to generate optical singularities. Optical singularities can be used for studying complex light fields, advancing our understanding of singular optics and light-matter interaction [85]. They also find applications in quantum communication and optical tweezers, and enable precise manipulation of microscopic particles, with further implications in fields such as optofluidic and biophotonics [86, 87].

For this study, we printed chiral nematic LC droplets (E7 + 4 wt.% R811, pitch = 2.5 µm) on a planar alignment layer, which generates optical singularities in polarization patterns. As shown in **Fig. 4a**, the POM image **(i)** displays a clear multi-ring spiral pattern, while the corresponding polarization properties are presented in **(ii)**. The fast axis orientation map reveals alternating black and white patterns, corresponding to the alignment of circular polarization state fast axis. The retardance map further highlights concentric regions with significant retardance effects, indicating that the droplet induces measurable phase delays for circularly polarized light in these areas. These observations suggest that the droplet exhibits localized characteristics of a circular retarder. The black-white transitions at the edge of the fast axis map are due to numerical discontinuities, as previously described.

As an example, **Fig.4b(i)** presents the polarization pattern observed when horizontally polarized light is used as the incident light, this pattern reveals the presence of optical singularities. The red curves in **Fig.4b(i)** highlights the L line spirals, representing regions where the polarization is linear, but the handedness is undefined [88]. Subsequently, **Fig.4b(ii)** shows the interference pattern generated by this printed chiral nematic

LC droplet, and the pattern displays multiple concentric rings, resulting from aberrated phase shift across the droplet.

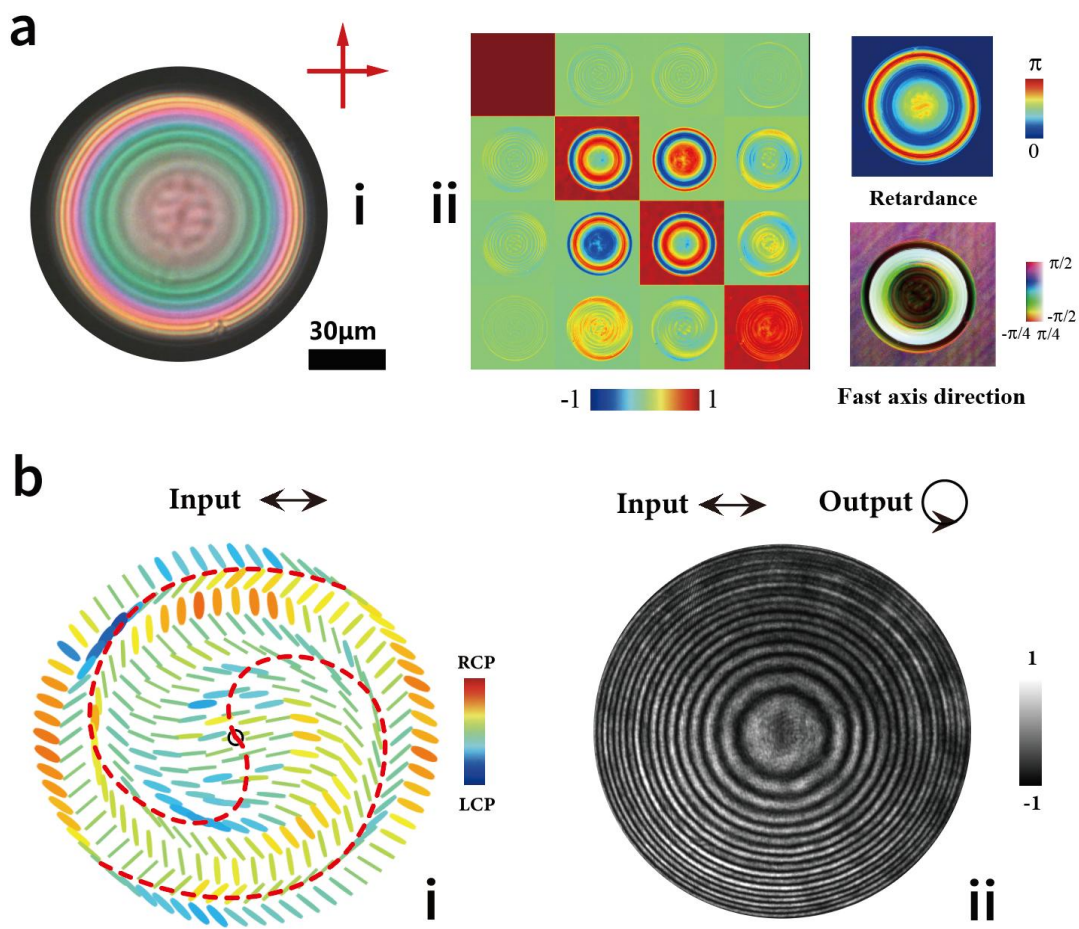

**Figure 4 Experimental results demonstrating the modulation of light by a printed chiral nematic LC droplet (E7 + 4 wt.% R811) on a glass substrate coated with a planar alignment layer.** a(i) Polarizing optical microscopy image of the printed droplet. The geometric properties of the droplets are as follows: diameter = 122 μm, height = 28 μm, and contact angle = 32.2°. The single-headed red arrows represent the orientations of the polarizers. a(ii) The MM components characterizing the droplet's full optical response with the retardance and fast axis direction profiles. b(i) The polarization pattern highlights optical singularities, with the red dashed lines representing L lines in the pattern. b(ii) The interference pattern shows concentric rings, resulting from radial variations in phase retardation as light passes through the droplet. Measurements were taken at a temperature of 23℃.

**Discussion**

This work proposes an innovative method for generating complex structured light fields using inkjet-printed LC droplets. By leveraging nematic LC droplets on homeotropic and planar alignment layers, as well as chiral nematic LC droplets on planar alignment layers, this approach exploits the intrinsic birefringence of LCs to expand the scope of structured light generations. Specifically, we demonstrate the generation of structured light fields, including skyrmionic structure with two units of OAM, azimuthal/radial polarized beams, and beam with polarization singularities.

Compared to conventional methods, printed LC droplets present distinct advantages in three key aspects. First, the inkjet-printing process is not only cost-effective but also highly efficient, capable of producing nearly a hundred droplets per second, enabling large-scale array fabrication. Second, the small size of LC droplets makes them ideal for miniaturized optical devices. Their compact footprint thus allows them to be easily integrated into applications where space is limited, such as on photonic chips. Third, their versatility allows for flexible design and precise tailoring of light fields without the need for complex optical assemblies. Such a flexibility is especially valuable in integrated photonics and optical sensing, where reducing system complexity and device footprint is critical.

A key challenge in generating diverse light patterns is that conventional alignment layers impose only a single, uniform director alignment (either planar or homeotropic), which limits the range of structured light that can be produced. Direct laser writing (DLW) on photoalignment layers doped with SD1 - a photosensitive azo-dye material - enables precise patterning of the alignment layer. Under laser exposure, the SD1 molecules reorient locally, allowing the creation of spatially varying director alignments [89]. In addition, the intrinsic sensitivity of LCs to electric fields enables real-time tuning of the droplet's optical response. By applying an electric field, the LC director configuration could be adjusted on demand, resulting in structured beams whose properties can be dynamically controlled. Furthermore, combining local curing with electrical tuning offers a promising strategy for controlling the LC droplet's optical response. In this approach, DLW is used to cure selected regions of the LC droplet, forming polymer networks that lock the LC director in place, while the remaining uncured regions remain responsive to electric fields. This strategy provides stable spatial control along with dynamic tunability, which is advantageous for generating structured light [90].

In summary, our study demonstrates that printed LC droplets can serve as effective passive elements for generating structured light. We believe this approach lays a solid foundation for future studies into the design and control of structured light fields through LC droplet systems. We hope this work inspires further developments across a range of disciplines, including integrated photonics, optical communication, quantum technologies, and biomedical imaging.

## Methods

*Material Preparation*

For all printed droplets, we used the well-characterized nematic LC mixture, E7 (Synthon Chemicals Ltd), composed of four compounds: 4-cyano-4′4′-n-pentyl-biphenyl (5CB), 4-cyano-4′4′-nheptyl-biphenyl (7CB), 4-cyano-4′4′-n-octyloxy-biphenyl, and 4-cyano-4″4″-n-pentyl-p-terphenyl. It possesses a moderate birefringence, $\Delta n$, of around 0.2 at visible wavelengths at room temperature, and a positive dielectric anisotropy [91]. For the chiral nematic mixture, E7 was dispersed with the low twisting power chiral dopant R811 ((R)-2-Octyl 4-[4-(Hexyloxy)benzoyloxy] benzoate; Merck KGaA) at various concentrations by weight.

*Substrate Preparation*

To promote a homeotropic alignment, forcing the LC director to align perpendicular to the glass substrate, a solution consisting of 0.02 wt.% lecithin in isopropyl alcohol was spin-coated onto the substrate at a speed of 1200 revolutions per second, then baked at 70°C for 10 seconds. For the planar alignment layers, a polyvinyl alcohol (PVA) solution was spin-coated onto the glass substrates at the same spin rate. After spin-coating, the PVA layer was baked at 70°C for 30 seconds to ensure complete drying. The substrates were then mechanically rubbed with a specialized cloth to provide a preferential alignment direction.

*Inkjet Printing Process*

The DoD inkjet printing process was carried out using MicroFab Jetlab II system (MicroFab Technologies) equipped with a MJ-AT-01-080-8MX printhead , featuring an 80 μm nozzle diameter. The printing process is computer-controlled, and the system's flexibility allows for the printing of arbitrary shapes with high precision. The actuation of the piezoelectric nozzle was driven by a carefully-designed trapezoidal voltage waveform, which allowed the generation of clean, satellite-free, droplets. The LC droplets were printed in a two-dimensional array onto the treated glass substrates with an inter-droplet separation of 150 μm

The printhead was operated at a temperature of 70°C, which is above the clearing point ($T_c$) of the nematic LC mixture E7 ($T_c$ = 58°C). At this temperature, the LC remains in the isotropic phase, ensuring a viscosity low enough for effective printing. Through a process of characterization, the optimum conditions of voltage waveform for driving the printhead were found to be as follows: a dwell voltage of 55 V, an echo voltage of -45 V, a rise time of 5 μs, a fall time of 5 μs, a dwell time of 15 μs, and an echo time of 10 μs. This precise control allows for the generation of uniform droplets for each pulse. Each droplet, with an initial in-flight diameter of approximately 80 μm and a volume of around 180 picolitres, retained a spherical shape as it moved through the air. To manufacture larger diameter droplets, a method of precisely depositing multiple droplets on the same location could be used to control the size of a single droplet [55].

**Acknowledgements**

ACJO acknowledges the Engineering and Physical Sciences Research Council (EPSRC-UK) for a graduate studentship (EP/R513295/1) and New College, Oxford for a research allowance. AACP acknowledges funding from the NSF/CBET-EPSRC (Grant Nos. EP/W016036/1 and EP/S029966/1). and the John Fell Fund via a Pump-Priming grant (0005176). WK, SJE and SMM acknowledge the EPSRC-UK for a research grant (EP/W022567/1).

**References**

[1] T. Bell, B. Li, S. Zhang, in *Wiley Encyclopedia of Electrical and Electronics Engineering*, **2016**.
[2] S. N. Khonina, A. V. Ustinov, A. P. Porfirev, *Photonics* **2023**, 10.
[3] O. V. Angelsky, A. Y. Bekshaev, S. G. Hanson, C. Y. Zenkova, I. I. Mokhun, J. Zheng, *Frontiers in Physics* **2020**, 8.
[4] Q. Zhan, *Advances in Optics and Photonics* **2009**, 1.
[5] T. Bauer, P. Banzer, E. Karimi, S. Orlov, A. Rubano, L. Marrucci, E. Santamato, R. W. Boyd, G. Leuchs, *Science* **2015**, 347, 964.
[6] C. Rosales-Guzmán, B. Ndagano, A. Forbes, *Journal of Optics* **2018**, 20.
[7] C. Maurer, A. Jesacher, S. Fürhapter, S. Bernet, M. Ritsch-Marte, *New Journal of Physics* **2007**, 9, 78.
[8] V. Parigi, V. D'Ambrosio, C. Arnold, L. Marrucci, F. Sciarrino, J. Laurat, *Nat Commun* **2015**, 6, 7706.
[9] M. J. Padgett, *Opt Express* **2017**, 25, 11265.
[10] A. M. Yao, M. J. Padgett, *Advances in Optics and Photonics* **2011**, 3.
[11] M. Li, S. J. Elston, C. He, X. Qiu, A. A. Castrejón-Pita, S. M. Morris, *Advanced Optical Materials* **2024**, 12.
[12] J. Courtial, D. A. Robertson, K. Dholakia, L. Allen, M. J. Padgett, *Physical Review Letters* **1998**, 81, 4828.
[13] G. Molina-Terriza, J. P. Torres, L. Torner, *Phys Rev Lett* **2002**, 88, 013601.
[14] M. Ritsch-Marte, *Philos Trans A Math Phys Eng Sci* **2017**, 375.
[15] S. W. Hell, J. Wichmann, *Opt. Lett.* **1994**, 19, 780.
[16] M. E. J. Friese, T. A. Nieminen, N. R. Heckenberg, H. Rubinsztein-Dunlop, *Nature* **1998**, 394, 348.
[17] V. Garces-Chavez, D. McGloin, M. J. Padgett, W. Dultz, H. Schmitzer, K. Dholakia, *Phys Rev Lett* **2003**, 91, 093602.
[18] M. Padgett, R. Bowman, *Nature Photonics* **2011**, 5, 343.
[19] C. He, J. Lin, J. Chang, J. Antonello, B. Dai, J. Wang, J. Cui, J. Qi, M. Wu, D. S. Elson, P. Xi, A. Forbes, M. J. Booth, *Optica* **2022**, 9, 1109.
[20] C. He, J. Chang, P. S. Salter, Y. Shen, B. Dai, P. Li, Y. Jin, T. Samlan Chandran, M. Li, T. Aziz, J. Wang, J. Antonello, D. Yang, Q. Ji, J. Lin, D. S. Elson, M. Zhang, H. He, H. Ma, M. J. Booth, *Advanced Photonics* **2022**, 4, 26001.
[21] C. He, M. Booth, *Opt. Photon. News* **2020**, 31, 47.
[22] C. He, J. Lin, B. Dai, P. Xi, M. Booth, presented at *Proc.SPIE*, **2020**.
[23] C. He, J. Chang, H. He, S. Liu, D. S. Elson, H. Ma, M. J. Booth, presented at *Proc.SPIE*, **2020**.
[24] F. Yue, D. Wen, J. Xin, B. D. Gerardot, J. Li, X. Chen, *ACS Photonics* **2016**, 3, 1558.

[25] V. D'Ambrosio, G. Carvacho, F. Graffitti, C. Vitelli, B. Piccirillo, L. Marrucci, F. Sciarrino, *Physical Review A* **2016**, 94.
[26] N. M. Litchinitser, *Science* **2012**, 337, 1054.
[27] Z.-X. Li, Y.-P. Ruan, P. Chen, J. Tang, W. Hu, K.-Y. Xia, Y.-Q. Lu, *Chin. Opt. Lett.* **2021**, 19, 112601.
[28] T. Giordani, A. Suprano, E. Polino, F. Acanfora, L. Innocenti, A. Ferraro, M. Paternostro, N. Spagnolo, F. Sciarrino, *Physical Review Letters* **2020**, 124, 160401.
[29] N. Radwell, R. D. Hawley, J. B. Götte, S. Franke-Arnold, *Nature Communications* **2016**, 7, 10564.
[30] Z.-Y. Rong, Y.-J. Han, S.-Z. Wang, C.-S. Guo, *Opt. Express* **2014**, 22, 1636.
[31] M.-Q. Cai, Z.-X. Wang, J. Liang, Y.-K. Wang, X.-Z. Gao, Y. Li, C. Tu, H.-T. Wang, *Appl. Opt.* **2017**, 56, 6175.
[32] Y. Zhang, P. Li, C. Ma, S. Liu, H. Cheng, L. Han, J. Zhao, *Appl. Opt.* **2017**, 56, 4956.
[33] Q. Hu, Y. Dai, C. He, M. J. Booth, *Optics Communications* **2020**, 459, 125028.
[34] A. S. Rao, P. Kumar, T. Omatsu, presented at *2023 Conference on Lasers and Electro-Optics Europe & European Quantum Electronics Conference (CLEO/Europe-EQEC)*, 26-30 June 2023, **2023**.
[35] M. Szatkowski, J. Masajada, I. Augustyniak, K. Nowacka, *Optics Communications* **2020**, 463, 125341.
[36] Q. Zhang, M. Gu, *Light: Science & Applications* **2024**, 13, 32.
[37] C. He, Q. Hu, Y. Dai, M. J. Booth, presented at *Imaging and Applied Optics Congress*, Washington, DC, 2020/06/22, **2020**.
[38] Q. Hu, C. He, M. J. Booth, *Journal of Optics* **2021**, 23, 065602.
[39] Y. Dai, C. He, J. Wang, R. Turcotte, L. Fish, M. Wincott, Q. Hu, M. J. Booth, *Opt. Express* **2019**, 27, 35797.
[40] H. Mei-Li, C. Mao-Ling, C. Chau-Jern, *Optical Engineering* **2007**, 46, 070501.
[41] F. Cardano, E. Karimi, S. Slussarenko, L. Marrucci, C. de Lisio, E. Santamato, *Appl. Opt.* **2012**, 51, C1.
[42] A. Rubano, F. Cardano, B. Piccirillo, L. Marrucci, *J. Opt. Soc. Am. B* **2019**, 36, D70.
[43] Y. Zhang, J. Gao, X. Yang, *Scientific Reports* **2019**, 9, 9969.
[44] Y. Bao, J. Ni, C.-W. Qiu, *Advanced Materials* **2020**, 32, 1905659.
[45] Y. Shen, C. He, Z. Song, B. Chen, H. He, Y. Ma, J. A. J. Fells, S. J. Elston, S. M. Morris, M. J. Booth, A. Forbes, *Physical Review Applied* **2024**, 21, 024025.
[46] C. He, J. Chang, Q. Hu, J. Wang, J. Antonello, H. He, S. Liu, J. Lin, B. Dai, D. S. Elson, P. Xi, H. Ma, M. J. Booth, *Nature Communications* **2019**, 10, 4264.
[47] A. A. Wang, Y. Ma, Y. Zhang, Z. Zhao, Y. Cai, X. Qiu, B. Dong, C. He,   **2024**.
[48] E. Parry, S. Bolis, S. J. Elston, A. A. Castrejón-Pita, S. M. Morris, *Advanced Engineering Materials* **2018**, 20, 1700774.
[49] J. Jiang, X. Chen, Z. Mei, H. Chen, J. Chen, X. Wang, S. Li, R. Zhang, G. Zheng, W. Li, *Micromachines* **2024**, 15.
[50] W. Kamal, M. Li, J.-D. Lin, E. Parry, Y. Jin, S. J. Elston, A. A. Castrejón-Pita, S. M. Morris, *Advanced Optical Materials* **2022**, 10, 2101748.
[51] J.-U. Park, M. Hardy, S. J. Kang, K. Barton, K. Adair, D. k. Mukhopadhyay, C. Y. Lee, M. S. Strano, A. G. Alleyne, J. G. Georgiadis, P. M. Ferreira, J. A. Rogers, *Nature Materials* **2007**, 6, 782.
[52] J.-U. Park, J. H. Lee, U. Paik, Y. Lu, J. A. Rogers, *Nano Letters* **2008**, 8, 4210.
[53] S. Mishra, K. L. Barton, A. G. Alleyne, P. M. Ferreira, J. A. Rogers, *Journal of Micromechanics and Microengineering* **2010**, 20, 095026.
[54] E. Parry, D.-J. Kim, A. A. Castrejón-Pita, S. J. Elston, S. M. Morris, *Optical Materials* **2018**, 80, 71.
[55] W. Kamal, J.-D. Lin, S. J. Elston, T. Ali, A. A. Castrejón-Pita, S. M. Morris, *Advanced Materials Interfaces* **2020**, 7, 2000578.
[56] A. M. Beckley, T. G. Brown, M. A. Alonso, *Opt. Express* **2010**, 18, 10777.
[57] C. He, J. Lin, J. Chang, J. Antonello, B. Dai, J. Wang, J. Cui, J. Qi, M. Wu, D. S. Elson, *Optica* **2022**, 9, 1109.
[58] S. Gao, F. C. Speirits, F. Castellucci, S. Franke-Arnold, S. M. Barnett, J. B. Götte, *Physical Review A* **2020**, 102, 053513.

[59] C. He, Y. Shen, A. Forbes, *Light: Science & Applications* **2022**, 11, 205.
[60] Y. Shen, Q. Zhang, P. Shi, L. Du, X. Yuan, A. V. Zayats, *Nature Photonics* **2024**, 18, 15.
[61] A. A. Wang, Z. Zhao, Y. Ma, Y. Cai, S. Morris, H. He, L. Luo, Z. Xie, P. Shi, Y. Shen, *arXiv preprint arXiv:2409.17390* **2024**.
[62] C. He, B. Chen, Z. Song, Z. Zhao, Y. Ma, H. He, L. Luo, T. Marozsak, A. Wang, R. Xu, *arXiv preprint arXiv:2311.18148* **2023**.
[63] A. A. Wang, Z. Zhao, Y. Ma, Y. Cai, R. Zhang, X. Shang, Y. Zhang, J. Qin, Z.-K. Pong, T. Marozsák, B. Chen, H. He, L. Luo, M. J. Booth, S. J. Elston, S. M. Morris, C. He, *Light: Science & Applications* **2024**, 13, 314.
[64] C. He, B. Chen, Z. Song, Z. Zhao, Y. Ma, H. He, L. Luo, T. Marozsak, A. A. Wang, R. Xu, P. Huang, J. Li, X. Qiu, Y. Zhang, B. Sun, J. Cui, Y. Cai, Y. Zhang, A. Wang, M. Wang, P. Salter, J. A. J. Fells, B. Dai, S. Liu, L. Guo, Y. He, H. Ma, D. J. Royston, S. J. Elston, Q. Zhan, C. Qiu, S. M. Morris, M. J. Booth, A. Forbes, *Nature Communications* **2025**, 16, 4902.
[65] A. C. J. Orr, X. Qiu, W. Kamal, T. C. Sykes, S. J. Elston, J. M. Yeomans, S. M. Morris, A. A. Castrejon-Pita, *Soft Matter* **2024**, 20, 7493.
[66] C. He, H. He, J. Chang, B. Chen, H. Ma, M. J. Booth, *Light: Science & Applications* **2021**, 10, 194.
[67] S.-Y. Lu, R. A. Chipman, *J. Opt. Soc. Am. A* **1996**, 13, 1106.
[68] A. C. Russell, L. Shih-Yau, presented at *Proc.SPIE*, **1997**.
[69] T. Xuan, H. Zhai, H. He, C. He, S. Liu, H. Ma, *Opt. Lett.* **2022**, 47, 5797.
[70] Y. Shi, C. Chen, L. Deng, N. Zeng, H. Li, Z. Liu, H. He, C. He, H. Ma, *Opt. Lett.* **2024**, 49, 3356.
[71] R. Hao, N. Zeng, Z. Zhang, H. He, C. He, H. Ma, *Opt. Lett.* **2024**, 49, 2273.
[72] J. Fan, N. Zeng, H. He, C. He, S. Liu, H. Ma, *Journal of Innovative Optical Health Sciences* **2024**, 18, 2343003.
[73] Z. Zhang, R. Hao, C. Shao, C. Mi, H. He, C. He, E. Du, S. Liu, J. Wu, H. Ma, *Opt. Lett.* **2023**, 48, 6136.
[74] Z. Zheng, S. Conghui, H. Honghui, H. Chao, L. Shaoxiong, M. Hui, *Journal of Biomedical Optics* **2023**, 28, 102905.
[75] Y. Shi, Y. Sun, R. Huang, Y. Zhou, H. Zhai, Z. Fan, Z. Ou, P. Huang, H. He, C. He, Y. Wang, H. Ma, *Frontiers in Physics* **2022**, Volume 10 - 2022.
[76] C. Jintao, H. Honghui, H. Chao, M. Hui, presented at *Proc.SPIE*, **2016**.
[77] R. Hao, N. Zeng, Z. Zhang, H. He, C. He, H. Ma, *Opt. Express* **2024**, 32, 3804.
[78] L. Deng, Z. Fan, B. Chen, H. Zhai, H. He, C. He, Y. Sun, Y. Wang, H. Ma, *International Journal of Molecular Sciences* **2023**, 24.
[79] Y. Jin, N. P. Spiller, C. He, G. Faulkner, M. J. Booth, S. J. Elston, S. M. Morris, *Light: Science & Applications* **2023**, 12, 242.
[80] C. Shao, B. Chen, H. He, C. He, Y. Shen, H. Zhai, H. Ma, *Frontiers in Chemistry* **2022**, Volume 10 - 2022.
[81] R. Zhang, X. Qiu, Y. Ma, Z. Zhao, A. A. Wang, J. Guo, J. Qin, S. J. Elston, S. M. Morris, C. He, *arXiv preprint arXiv:2505.19811* **2025**.
[82] Y. Xue, C. Kuang, S. Li, Z. Gu, X. Liu, *Opt. Express* **2012**, 20, 17653.
[83] C. J. R. Sheppard, A. Choudhury, *Appl. Opt.* **2004**, 43, 4322.
[84] F. Tang, Y. Wang, L. Qiu, W. Zhao, Y. Sun, *Appl. Opt.* **2014**, 53, 7407.
[85] F. Cardano, E. Karimi, L. Marrucci, C. de Lisio, E. Santamato, *Opt. Express* **2013**, 21, 8815.
[86] I. Freund, *Optics Communications* **2002**, 201, 251.
[87] C. Alpmann, C. Schlickriede, E. Otte, C. Denz, *Scientific Reports* **2017**, 7, 8076.
[88] H. Rubinsztein-Dunlop, A. Forbes, M. V. Berry, M. R. Dennis, D. L. Andrews, M. Mansuripur, C. Denz, C. Alpmann, P. Banzer, T. Bauer, E. Karimi, L. Marrucci, M. Padgett, M. Ritsch-Marte, N. M. Litchinitser, N. P. Bigelow, C. Rosales-Guzmán, A. Belmonte, J. P. Torres, T. W. Neely, M. Baker, R. Gordon, A. B. Stilgoe, J. Romero, A. G. White, R. Fickler, A. E. Willner, G. Xie, B. McMorran, A. M. Weiner, *Journal of Optics* **2017**, 19, 013001.
[89] Y. Shi, P. S. Salter, M. Li, R. A. Taylor, S. J. Elston, S. M. Morris, D. D. C. Bradley, *Advanced Functional Materials* **2021**, 31, 2007493.

[90] A. Xu, C. Nourshargh, P. S. Salter, C. He, S. J. Elston, M. J. Booth, S. M. Morris, *ACS Photonics* **2023**, 10, 3401.
[91] J. Li, G. Baird, Y.-H. Lin, H. Ren, S.-T. Wu, *Journal of the Society for Information Display* **2005**, 13, 1017.